\documentclass[aps,prd,reprint,superscriptaddress,floatfix,nofootinbib]{revtex4-2}
\usepackage{amsfonts,amsmath,amssymb,bm,amsthm}
\usepackage{fnpct} 
\usepackage{comment}
\usepackage{ifpdf}
\usepackage{graphicx}
\usepackage{slashed}
\usepackage{color}
\usepackage[mathscr]{eucal}
\usepackage[utf8]{inputenc}
\usepackage{physics}
\usepackage{cancel}
\usepackage{soul}
\usepackage{booktabs}
\usepackage{simpler-wick}
\usepackage{hyperref}
\usepackage{tensor}
\usepackage{upgreek}
\usepackage{tikz}

\hypersetup{hidelinks}\def\Eprint#1#2{\href{#1}{\nolinkurl{#2}}}
\begin{document}

\title{Gravitational waves from cosmological first-order phase transitions with fluctuation-induced friction}

\author{Jun-Ye Chen}
\email{chenjunye@itp.ac.cn}
\affiliation{Institute of Theoretical Physics, Chinese Academy of Sciences, P.O. Box 2735, Beijing 100190, China}
\affiliation{School of Physical Sciences, University of Chinese Academy of Sciences, No.19A Yuquan Road, Beijing 100049, China}

\author{Dongdong Wei}
\email{weidd@ucas.ac.cn}
\affiliation{School of Fundamental Physics and Mathematical Sciences, Hangzhou Institute for Advanced Study, University of Chinese Academy of Sciences, Hangzhou 310024, China}
\affiliation{School of Physical Sciences, University of Chinese Academy of Sciences, No.19A Yuquan Road, Beijing 100049, China}
\affiliation{Institute of Theoretical Physics, Chinese Academy of Sciences, P.O. Box 2735, Beijing 100190, China}

\author{Zong-Kuan Guo}
\email{guozk@itp.ac.cn}

\affiliation{Institute of Theoretical Physics, Chinese Academy of Sciences, P.O. Box 2735, Beijing 100190, China}
\affiliation{School of Physical Sciences, University of Chinese Academy of Sciences, No.19A Yuquan Road, Beijing 100049, China}
\affiliation{School of Fundamental Physics and Mathematical Sciences, Hangzhou Institute for Advanced Study, University of Chinese Academy of Sciences, Hangzhou 310024, China}

\begin{abstract}

We study gravitational-wave production in a cosmological first-order phase transition with fluctuation-induced friction using three-dimensional lattice simulations of two coupled scalar fields. Kinetic and gradient energy in the additional field \(s\) accumulates near the expanding bubble walls of \(\phi\) and affects their motion. The contribution from \(s\) exceeds that from \(\phi\) at the spectral peak and has a slope similar to that of the bulk flow model over a finite interval above the peak. In the high-wave-number (UV) direction, \(\phi\) dominates and the total spectrum develops a broad feature. For the larger coupling, slowly contracting false vacuum remnants persist long after bubble merger.

\end{abstract}

\maketitle

\section{Introduction}

Cosmological first-order phase transitions (FOPTs) provide a connection between particle physics and gravitational-wave observations. They occur in many extensions of the Standard Model and proceed through the nucleation, expansion and collision of bubbles of a new phase \cite{Caprini:2015zlo,Caprini:2019egz,Profumo:2014opa,Dorsch:2016nrg,Coleman:1977py,Linde:1981zj}. The resulting stresses can generate a stochastic gravitational-wave background (SGWB). Its amplitude and spectral shape depend on the dynamics of the transition and can therefore carry information about the underlying theory. Space-based interferometers such as LISA, Taiji and TianQin are designed to detect gravitational waves in the millihertz frequency range \cite{LISA:2017pwj,Luo:2021qji,TianQin:2015yph}. A sufficiently strong first-order phase transition could produce a background detectable by LISA \cite{Caprini:2015zlo,Caprini:2019egz}. Predictions for these signals require a description of the sources during bubble growth, collision and the subsequent evolution.

The energy released during a transition can remain in scalar-field gradients or be transferred to other degrees of freedom. In a plasma, it can also contribute to heating and bulk fluid motion. The relative importance of these contributions depends on the interaction of the bubble walls with the surrounding medium. Friction and hydrodynamic backreaction can limit the wall velocity, changing how much energy is available in each source \cite{Huber:2013kj,Hindmarsh:2015qta}. The gravitational-wave spectrum depends not only on this energy distribution, but also on the spatial structure of the stresses and the time over which they remain active. Bubble-wall motion and the evolution after collision therefore both affect the gravitational-wave power spectrum.

The envelope approximation gives a simple description of gravitational-wave production from bubble collisions. It represents the source by thin walls and includes only their uncollided parts. The source disappears where bubbles overlap \cite{Kosowsky:1992vn,Huber:2008hg}. This approximation relates the peak of the spectrum to the typical bubble separation, but does not describe the subsequent evolution of the scalar field. Lattice simulations of vacuum transitions show a steeper decrease of the spectrum at high wave number and an additional contribution from field oscillations after collision \cite{Cutting:2018tjt}. These results demonstrate the importance of following the field evolution beyond the collision stage. Extensions of the envelope approximation and the bulk flow model (BFM) also allow the source to continue after collision, with assumptions about how it propagates and decreases \cite{Jinno:2017fby,Konstandin:2017sat}.

The evolution is different when a substantial fraction of the released energy enters the surrounding plasma. Numerical simulations of a scalar field coupled to a relativistic fluid show that sound waves can continue to generate gravitational waves long after the bubbles have merged \cite{Hindmarsh:2013xza,Hindmarsh:2015qta}. The duration of this acoustic source can greatly affect the signal. The sound shell model describes the velocity field as a superposition of sound waves from individual expanding bubbles. The mean bubble separation and the sound shell width set two characteristic length scales \cite{Hindmarsh:2016lnk,Hindmarsh:2017gnf}. Nonlinear fluid evolution and turbulence can introduce further changes \cite{Kamionkowski:1993fg,Caprini:2018mtu}. These descriptions depend on the dynamics of a fluid and cannot be applied to an interacting scalar system on the basis of a similar spatial distribution alone.

An additional scalar field can affect both bubble-wall motion and the gravitational-wave source. A change in its effective mass across the wall can cause reflection of field modes, with energy transferred between the field and the moving wall. The resulting backreaction couples the evolution of these fluctuations to the expansion of the bubbles. Recent lattice studies have investigated this effect in models with two scalar fields. An initially fluctuating background can change the wall motion \cite{Wei:2026qph}. Energy transfer has also been studied for a field that acquires mass during the transition \cite{Li:2025zxa}.  In particular, a classical scalar field does not by itself imply a thermal fluid description. Its energy distribution and its contribution to gravitational waves must be determined from the field evolution.

These considerations lead us to study gravitational-wave production in the presence of fluctuation-induced friction. The energy near an expanding wall can contribute to the source while also changing the wall motion. When bubbles collide, this energy need not follow the propagation assumed in either the envelope approximation or an acoustic model. It may spread through the new phase or remain concentrated in regions of false vacuum. We therefore study the gravitational-wave contributions of both fields, follow their energy distribution after collision, and compare the source with a model of expanding shells. This comparison requires both the spatial form of the stress and its time dependence.

We address these questions with three-dimensional lattice simulations of the scalar order parameter \(\phi\) coupled to an additional scalar field \(s\). The field \(\phi\) undergoes the transition, while \(s\) is initialized with Gaussian fluctuations. Starting from the initial bubbles, both fields evolve through expansion and collision according to the coupled classical field equations, without a fluid description or a phenomenological damping term.  We study two sets of parameters and resolve the gravitational-wave contributions from both fields. A separate simulation follows the scalar fields for a longer time, without evolving gravitational waves, to examine the evolution of the remaining false vacuum.

The simulations show that kinetic and gradient energy is concentrated in shells near the expanding walls. At the peak of the total gravitational-wave spectrum, the contribution from \(s\) is larger than that from \(\phi\) for both parameter sets. Over the finite interval \(8\leq kR_*\leq20\), the spectrum from \(s\) has a slope close to that of the BFM reference curve. The thin shell analysis clarifies the assumptions behind this comparison.  In the high-wave-number (UV) direction, the contribution from \(\phi\) dominates and the total spectrum develops a broad feature. For the larger coupling, regions of false vacuum containing \(s\)-field energy remain long after bubble collisions. We discuss their subsequent contraction without assuming thermal equilibrium. Their final state remains open.

Section~\ref{Sec:Physical Model} introduces the model. Section~\ref{Sec:isolated} describes the initial conditions and single-bubble evolution, and Sec.~\ref{Sec:multi} presents the bubble collisions and subsequent energy transfer. Section~\ref{Sec:GW} gives the gravitational-wave spectra, followed by the conclusions in Sec.~\ref{Sec:Conclusion and Discussion}. The stress of a thin shell and its relation to the BFM are discussed in Appendix~\ref{app:A}.

\section{Physical Model}
\label{Sec:Physical Model}
We consider two real scalar fields: the order parameter \(\phi\) and an additional scalar field \(s\) coupled to it. Their phenomenological effective potential has the polynomial form of Ref.~\cite{Wei:2026qph}. The parameters are fixed during the evolution.
\begin{equation}
\label{eq:potential}
\begin{aligned}
    V(\phi, s) = 
    & \frac{1}{2}m_\phi^2 \phi^2 + \frac{1}{4}\lambda_\phi \phi^4 +  \frac{\mu}{2\sqrt{2}} \phi^3 \\
    +& \frac{1}{2}m_s^2 s^2 + \frac{1}{4}\lambda_s s^4 + \frac{1}{4}\lambda_{\phi s} \phi^2 s^2.
\end{aligned}
\end{equation}

For \(s=0\), the potential has a false vacuum at \(\phi=0\) and a true vacuum at \(\phi=\phi_{\rm true}>0\). 
For the parameters used here, the cubic term with \(\mu<0\) produces a barrier separating the metastable and stable minima.
The nonzero minimum is determined by
\begin{equation}
    \lambda_\phi \phi^2+\frac{3\mu}{2\sqrt{2}}\phi+m_\phi^2=0 ,
\end{equation}
where \(\phi_{\rm true}\) denotes the larger positive solution. 
The released vacuum energy density is defined as
\begin{equation}
    \rho_{\rm vac}=V(0,0)-V(\phi_{\rm true},0).
\end{equation}
Inside bubbles of the new phase, \(\phi\) approaches \(\phi_{\rm true}\). Bubble walls separate these regions from the surrounding false vacuum.

The self-interaction \(\lambda_s s^4/4\) controls the nonlinear evolution of the \(s\) field, while the interaction term \(\lambda_{\phi s}\phi^2s^2/4\) makes its mass scale depend on the local value of \(\phi\). 
At fixed \(\phi\), the effective mass squared is defined by the curvature of the potential at \(s=0\):
\begin{equation}
\label{eq:meff}
m_{s,\rm eff}^2(\mathbf{x},t)
\equiv
\left.\frac{\partial^2 V}{\partial s^2}\right|_{s=0}
=
m_s^2+\frac{1}{2}\lambda_{\phi s}\phi^2(\mathbf{x},t).
\end{equation}
In the false vacuum, where \(\phi=0\), this reduces to
\begin{equation}
    m_{s,\rm false}^2=m_s^2 .
\end{equation}
Inside the bubbles, where \(\phi\simeq \phi_{\rm true}\), it approaches
\begin{equation}
    m_{s,\rm true}^2
    =
    m_s^2+\frac{1}{2}\lambda_{\phi s}\phi_{\rm true}^2 .
\end{equation}
In the linear approximation, the change in mass across a locally planar wall suppresses the transmission of sufficiently low-frequency \(s\) excitations in the wall frame.

The field equations following from Eq.~\eqref{eq:potential} are
\begin{equation}\label{eq:eom_phi}
    \ddot\phi - \nabla^2\phi + m^2_\phi\phi + \lambda_\phi\phi^3 + \frac{3\mu}{2\sqrt 2}\phi^2 + \frac{1}{2}\lambda_{\phi s}s^2\phi = 0\,,
\end{equation}
\begin{equation} \label{eq:eom_s}
\ddot s - \nabla^2 s + m^2_s s + \lambda_s s^3 + \frac{1}{2}\lambda_{\phi s}\phi^2 s = 0\,. 
\end{equation}
The interaction terms in the field equations describe both the change in the mass of \(s\) and the backreaction on \(\phi\). We estimate the force that opposes expansion by projecting \(\lambda_{\phi s}s^2\phi/2\) onto the wall profile. For an approximately planar wall, choose the normal coordinate \(\xi\) to increase from the false vacuum side to the true vacuum side. Then \(\phi(-\infty)=0\), \(\phi(+\infty)=\phi_{\rm true}\), and \(\partial_\xi\phi>0\). The contribution that reduces the driving pressure is
\begin{equation}
\delta p_s
\simeq
\int_{\rm wall} d\xi\,
\frac{1}{2}\lambda_{\phi s}
s^2(\xi)\phi(\xi)\partial_\xi\phi(\xi).
\label{eq:delta-ps-wall-projection}
\end{equation}
Thus the effective pressure on the wall may be written schematically as
\begin{equation}
p_{\rm eff}
\simeq
\rho_{\rm vac}-\delta p_s .
\label{eq:effective-pressure}
\end{equation}
For \(\lambda_{\phi s}>0\) and a monotonic wall profile, \(\delta p_s\) is non-negative and reduces the outward driving pressure.
The weighted average across the wall is defined as
\begin{equation}
\langle s^2\rangle_{\rm wall}
\equiv
\frac{2}{\phi_{\rm true}^2}
\int_{\rm wall}d\xi\,
s^2(\xi)\phi(\xi)\partial_\xi\phi(\xi).
\label{eq:s2-wall-average}
\end{equation}
With this definition, Eq.~\eqref{eq:delta-ps-wall-projection}
becomes
\begin{equation}
\delta p_s
\simeq
\frac{1}{4}
\lambda_{\phi s}
\langle s^2\rangle_{\rm wall}
\phi_{\rm true}^2 .
\label{eq:delta-ps-estimate}
\end{equation}
The backreaction depends on \(s^2\) across the wall. For a fixed wall profile and coupling, a larger \(\langle s^2\rangle_{\rm wall}\) gives a larger \(\delta p_s\) in Eq.~\eqref{eq:delta-ps-estimate}, reducing the pressure that drives expansion.

In scalar--fluid descriptions, the wall interacts with a plasma through hydrodynamic backreaction and, in some treatments, a phenomenological damping term \cite{Huber:2013kj,Hindmarsh:2013xza}. Here the backreaction follows directly from the coupled classical field equations, without fluid variables or an assumed hydrodynamic equation of state \cite{Wei:2026qph}. Quantum particle production effects, including transition radiation at ultra-relativistic walls \cite{Bodeker:2017cim}, are also not included.

\section{Initial conditions and single-bubble evolution}
\label{Sec:isolated}

Single-bubble simulations give the wall motion for each parameter set before collisions begin. All \(\phi\) bubbles are initialized simultaneously at chosen positions and radii, with stochastic fluctuations only in \(s\). Alternatively, one can add fluctuations to both fields and let \(\phi\) bubbles nucleate during the evolution. In the examples considered, either initialization gives an \(s\) contribution comparable to or larger than the \(\phi\) contribution at the peak of the total GW spectrum. The resulting spectra differ from those of vacuum transitions with a single scalar field, as discussed in Sec.~\ref{Sec:GW}. Introducing bubbles in the initial state provides a controlled period of expansion before collision, allowing us to study the backreaction of \(s\) on the walls and subsequent GW production.

{
A spherically symmetric critical configuration at \(s=0\) provides a reference bubble profile. It satisfies the same \(O(3)\)-symmetric equation used for thermal critical bubbles \cite{Linde:1981zj}. 
}
\begin{align}
    \phi_b''(r)+\frac{2}{r}\phi_b'(r)
    &=
    \left.
    \partial_\phi V(\phi,s)
    \right|_{s=0,\ \phi=\phi_b},
    \label{eq:bounce}\\
    \phi_b'(0)&=0,
    \qquad
    \phi_b(r\rightarrow\infty)=0 .
    \label{eq:bounce_bc}
\end{align}
Here, \(r\) is the radial coordinate, and \(\phi_b'=d\phi_b/dr\).

{
The bounce determines the reference scales for the critical radius \(R_c\) and wall thickness \(l_w\). For each initial bubble, the field profile is
\begin{equation}
    \phi(r,t_0)
    =
    \frac{\phi_{\rm true}}{2}
    \left[1-\tanh\!\left(\frac{r-R_0}{l_w}\right)\right],
    \qquad
    \dot\phi(r,t_0)=0,
    \label{eq:lattice-bubble-profile}
\end{equation}
where \(t_0\) is the initial time of the simulations and a dot denotes a time derivative. 
Fluctuations in \(s\) can cause bubbles initialized at the critical radius \(R_c\) to collapse. 
We therefore choose \(R_0>R_c\) to make expansion more likely, although some initial contraction may occur. This choice does not change the qualitative physical picture.
}

{
During preheating after inflation, coherent inflaton oscillations can resonantly amplify bosonic modes and generate large nonthermal scalar fluctuations with high occupation numbers \cite{Kofman:1995fi,Greene:1997fu}. Once these modes become highly occupied, their subsequent evolution admits an approximately classical field description. Nonlinear lattice studies have further shown that such amplified fluctuations can significantly modify the effective potential and restore a symmetry nonthermally; in suitable models, the subsequent symmetry breaking proceeds through a first-order phase transition with bubble nucleation \cite{Khlebnikov:1998sz}. Motivated by this class of scenarios, we model the scalar field \(s\) at the onset of the transition as a classical Gaussian random field with a finite ultraviolet cutoff. We do not evolve a specific preheating model continuously from the end of inflation to the onset of the transition. Accordingly, the amplitude and spectral shape of the residual \(s\)-field fluctuations at the transition are treated phenomenologically. Our setup is therefore intended to study the subsequent response of an already formed bubble wall to a pre-existing nonthermal scalar background, rather than to provide a microscopic prediction for the origin or detailed spectrum of that background.
At the initial time \(t_0\), we use the Fourier convention
\begin{equation}
s(\mathbf{x},t_0)
=\int\frac{d^3k}{(2\pi)^3}\,
\tilde s(\mathbf{k})e^{i\mathbf{k}\cdot\mathbf{x}},
\end{equation}
and similarly for \(\pi_s=\dot{s}\). The initial Gaussian fluctuations are specified by
\begin{equation}
\begin{aligned}
\left\langle\tilde s(\mathbf{k})\tilde s^*(\mathbf{k}')\right\rangle
&=(2\pi)^3\delta^{(3)}(\mathbf{k}-\mathbf{k}')
\frac{A_s^2}{\omega_{\mathbf{k}}},\\
\left\langle\tilde\pi_s(\mathbf{k})\tilde\pi_s^*(\mathbf{k}')\right\rangle
&=(2\pi)^3\delta^{(3)}(\mathbf{k}-\mathbf{k}')
A_s^2\omega_{\mathbf{k}}.
\end{aligned}
\label{eq:initial-gaussian}
\end{equation}
Here \(\omega_{\mathbf{k}}^2=k^2+m_s^2\). The brackets denote an ensemble average. Independent modes are sampled with zero mean. The field--momentum covariance vanishes, and the remaining modes satisfy the reality conditions. Only modes with \(0<k\leq k_{\rm cut}\) are included in this definition. In the numerical initialization, the integral is evaluated as a sum over the modes of a periodic box. The quantity \(k^2\) is replaced by the eigenvalue of the negative lattice Laplacian. The cutoff is applied to the square root of this eigenvalue. The zero mode and modes above the cutoff are set to zero. The free parameter \(A_s\) controls the overall fluctuation amplitude independently of the parameters of the scalar potential. In the massless continuum limit, the field spectrum is proportional to \(1/k\) below the cutoff, as in Ref.~\cite{Li:2025zxa}.

}

\begin{figure}[!t]
    \centering
    \includegraphics[width=0.98\linewidth]{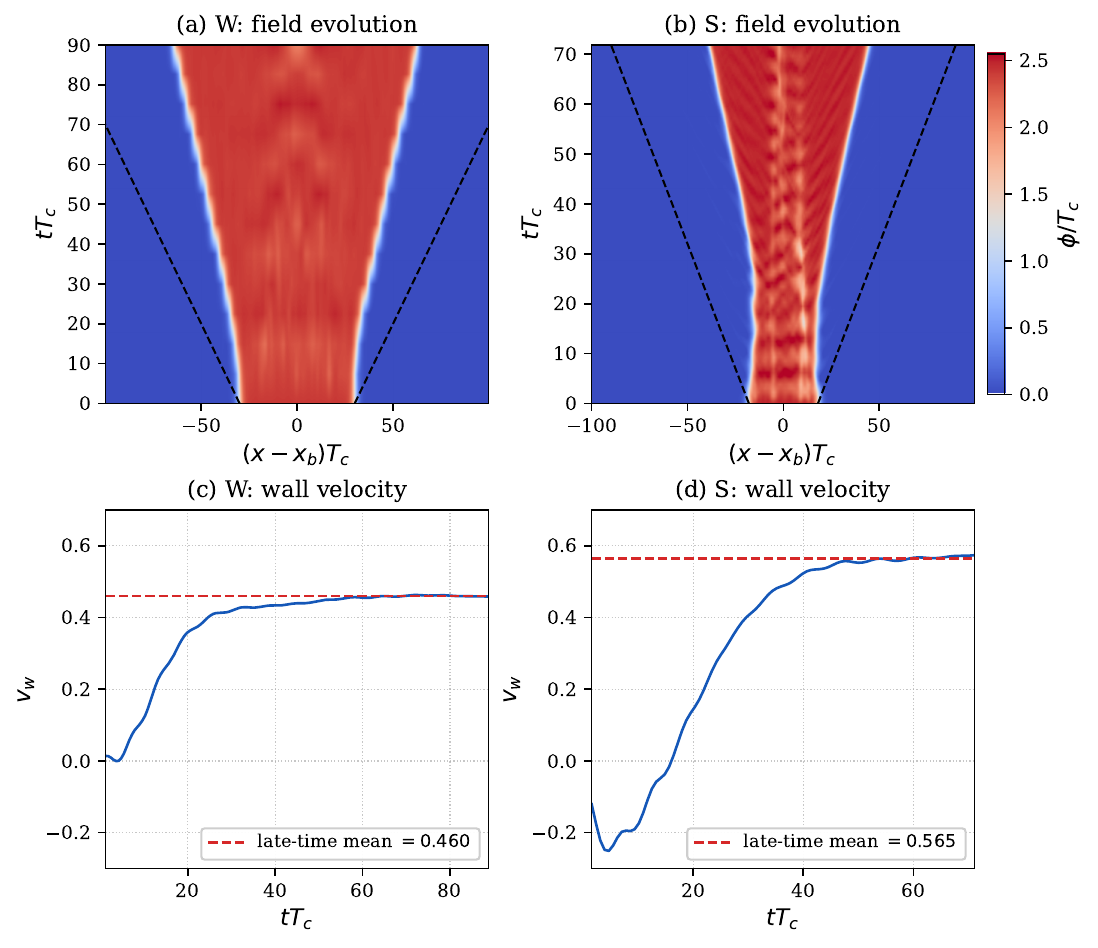}
    \caption{
    Single-bubble evolution for W (left) and S (right). The upper panels show \(\phi/T_c\) along a line through the bubble centre. For display, linear interpolation is used between the recorded times. Dashed curves show propagation from the initial wall at the speed of light. The lower panels show \(v_w=dR_b/dt\), smoothed over five samples, where \(R_b\) is the radius defined from the volume of the region with \(\phi\geq\phi_{\rm true}/2\). Horizontal dashed lines show the mean velocities over the final \(30\%\) of each run, \(\bar v_w=0.460\) and \(0.565\). 
    }
    \label{fig:Single_Bubble_Evolution}
\end{figure}

{

Parameters are expressed in units of the fixed reference scale \(T_c\). 
We study two representative parameter sets, W and S, for which the bubble wall velocities remain below the speed of light.
The parameters of the model and the initial conditions are listed in Table~\ref{tab:Parameters}. 
Fig.~\ref{fig:Single_Bubble_Evolution} shows the single-bubble evolution. The bubble volume \(V_b(t)\) is the volume with \(\phi\geq\phi_{\rm true}/2\). The corresponding bubble radius is \(R_b(t)=[3V_b(t)/(4\pi)]^{1/3}\). The velocity \(v_w=dR_b/dt\) is smoothed with a moving average over five samples. Its mean over the final \(30\%\) of each run is \(\bar v_w\simeq0.460\) for W and \(0.565\) for S.

}

\begin{table*}[!t]
\centering
\caption{
Simulation parameters. Here \(N_{\rm grid}^3\) is the number of lattice sites, \(\Delta x\) is the lattice spacing, and \(\Delta t\) is the time step. All setups use \(m_\phi/T_c=m_s/T_c=1\), \(\lambda_\phi=0.4\), \(\lambda_s=0.5\), \(l_wT_c=2.043\), \(m_\phi\Delta x=0.390625\), and \(\Delta t=0.2\Delta x\).
}
{\small
\setlength{\tabcolsep}{6pt}
\begin{tabular*}{\textwidth}{@{\extracolsep{\fill}}l*{8}{c}@{}}
\hline\hline
 & $\mu/T_c$ & $\lambda_{\phi s}$ & $A_s^2$ & $k_{\rm cut}/T_c$ & $R_0T_c$ & $m_\phi L$ & $N^3_{\rm grid}$ & $N_b$ \\
\hline
W & $-1.30$ & 2.0 & 0.61035 & 1.8616 & 30.0 & 400 & $1024^3$ & 32 \\
S & $-1.315$ & 7.5 & 0.56802 & 1.6776 & 18.0 & 400 & $1024^3$ & 32 \\
S$_L$ & $-1.315$ & 7.5 & 0.57019 & 1.6776 & 18.0 & 200 & $512^3$ & 4 \\
\hline\hline
\end{tabular*}
}
\label{tab:Parameters}
\end{table*}

\section{MULTI-BUBBLE MERGER AND EVOLUTION OF ENERGY SHELLS}
\label{Sec:multi}

\subsection{Bubble expansion and formation of energy shells}

The mean bubble separation is \(R_*=(L^3/N_b)^{1/3}\), where \(L^3\) is the box volume and \(N_b\) is the initial bubble number. For parameter sets W and S, the fields are evolved on \(1024^3\) lattices. Each simulation starts with \(N_b=32\) bubbles at \(t=t_0\). Their centres are chosen randomly with a minimum separation \(m_\phi r_{\rm min}=100.0\), calculated using periodic boundary conditions. This leaves the initial walls separated and allows a period of expansion before collision. The extended simulation S$_L$ uses \(512^3\) sites, four bubbles, and \(m_\phi L=200\), and follows the scalar fields to \(t/R_*=30\) without evolving gravitational waves. All three use the time step \(\Delta t=0.2\Delta x\) and are performed with a modified version of LATfield2~\cite{Daverio:2015ryl}.

Before collisions, the backreaction of \(s\) affects the motion of each wall, as in the single-bubble simulations. The kinetic and gradient energy of \(s\) is concentrated in shells near the expanding walls.  The energy shells in \(s\) then interact during bubble merger. After collisions, the walls and shells spread through the new phase in W, whereas localized \(s\)-field energy remains inside regions of false vacuum in S. Collisions distort the approximately spherical field profiles and redistribute energy among kinetic, gradient, and potential contributions. The resulting configurations are not well described by freely propagating energy shells in \(s\).

\begin{figure*}[t]
\centering
\includegraphics[width=0.95\textwidth]{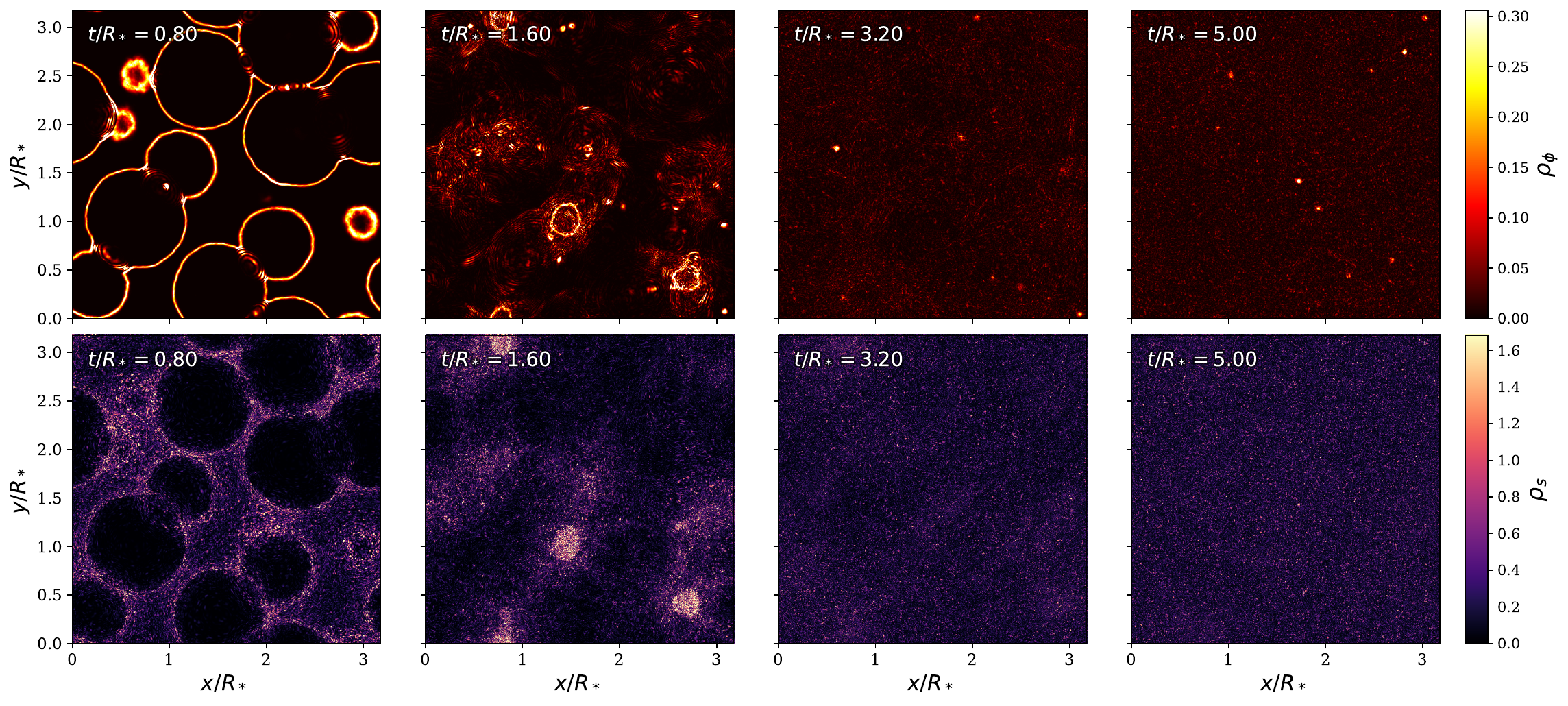}
\par\smallskip {\small (a) W}
\par\medskip
\includegraphics[width=0.95\textwidth]{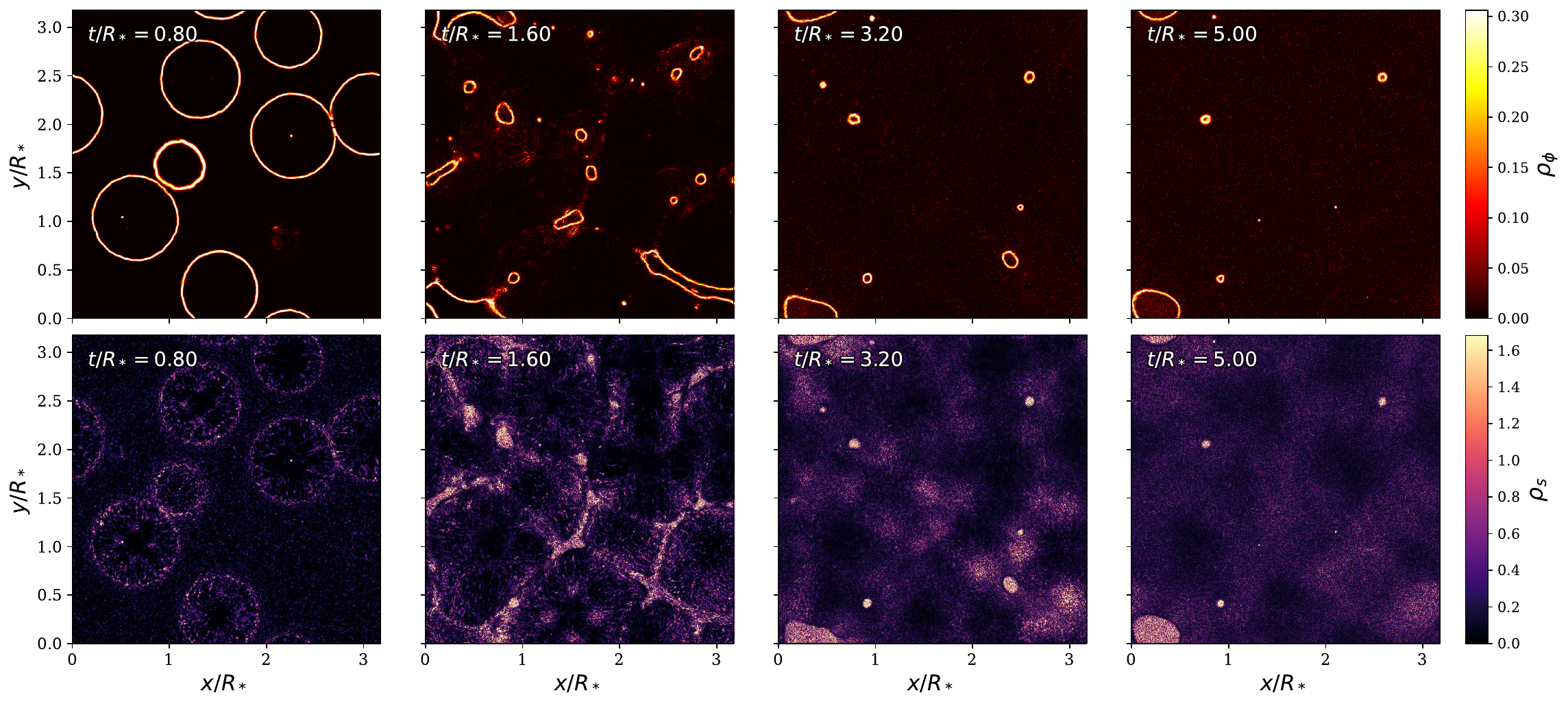}
\par\smallskip {\small (b) S}
\caption{The sum of the kinetic and gradient energy densities for (a) W and (b) S on the central \(z=L/2\) slice at \(t/R_*=0.80,\ 1.60,\ 3.20,\ 5.00\). Within each panel, the upper and lower rows show \(\rho_\phi=\rho_{\mathrm K,\phi}+\rho_{\mathrm G,\phi}\) and \(\rho_s=\rho_{\mathrm K,s}+\rho_{\mathrm G,s}\), respectively, as defined in Eq.~\eqref{eq:energy-densities}. Corresponding rows in (a) and (b) use the same colour scale. In W, the walls and shells have largely spread through the new phase by \(5R_*\). In S, localized \(s\)-field energy remains inside regions of false vacuum bounded by \(\phi\) walls through \(5R_*\), as seen on the plotted slice.}
\label{fig:WS_energy_density_evolution}
\end{figure*}

The local kinetic and gradient energy densities, for \(a=\phi,s\), are
\begin{equation}
    \rho_{\mathrm K,a}=\frac{1}{2}\dot a^2,
    \qquad
    \rho_{\mathrm G,a}=\frac{1}{2}(\nabla a)^2,
    \qquad
    \rho_a=\rho_{\mathrm K,a}+\rho_{\mathrm G,a}.
\label{eq:energy-densities}
\end{equation}
The sum \(\rho_a\) does not include potential energy. We further monitor the shifted potential energy density
\begin{equation}
    \rho_{\rm pot}=V(\phi,s)-V(\phi_{\rm true},0),
\end{equation}
together with the field configuration to identify candidate regions of false vacuum.
\subsection{Energy transfer and the evolution of false vacuum regions} \label{Subsec: False-Vacuum Droplets}
\begin{figure*}[t]
    \centering
    \includegraphics[width=0.98\textwidth]{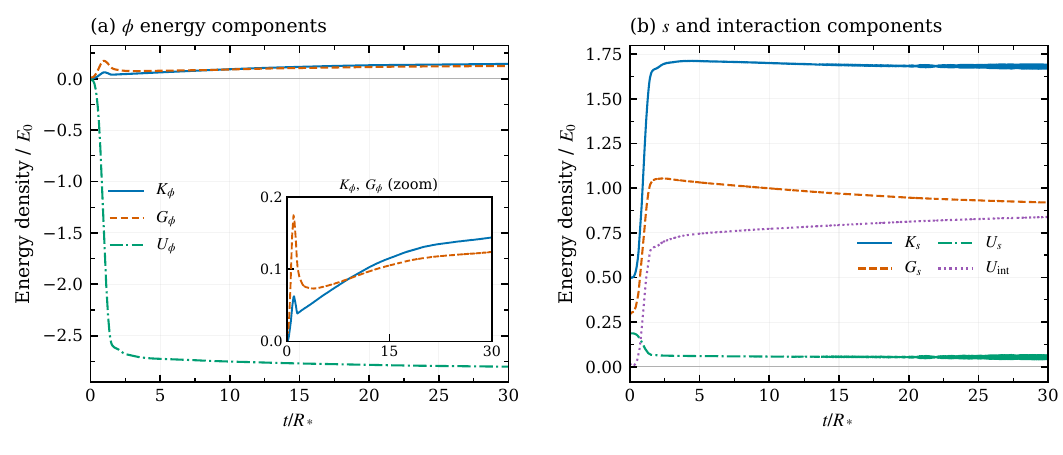}
    \caption{
    Evolution of the spatially averaged energy components in S$_L$. (a) Kinetic, gradient, and potential energies of \(\phi\). The inset in panel (a) shows the kinetic and gradient terms on a smaller scale. (b) The corresponding \(s\) components and the interaction energy \(U_{\rm int}\), shown separately. All components are normalized by the initial total energy density \(E_0\), with no shift in the potential. The curves are not smoothed.
    }
    \label{fig:2_energy_all_components}
\end{figure*}

\begin{figure*}[t]
    \centering
    \includegraphics[width=0.98\textwidth]{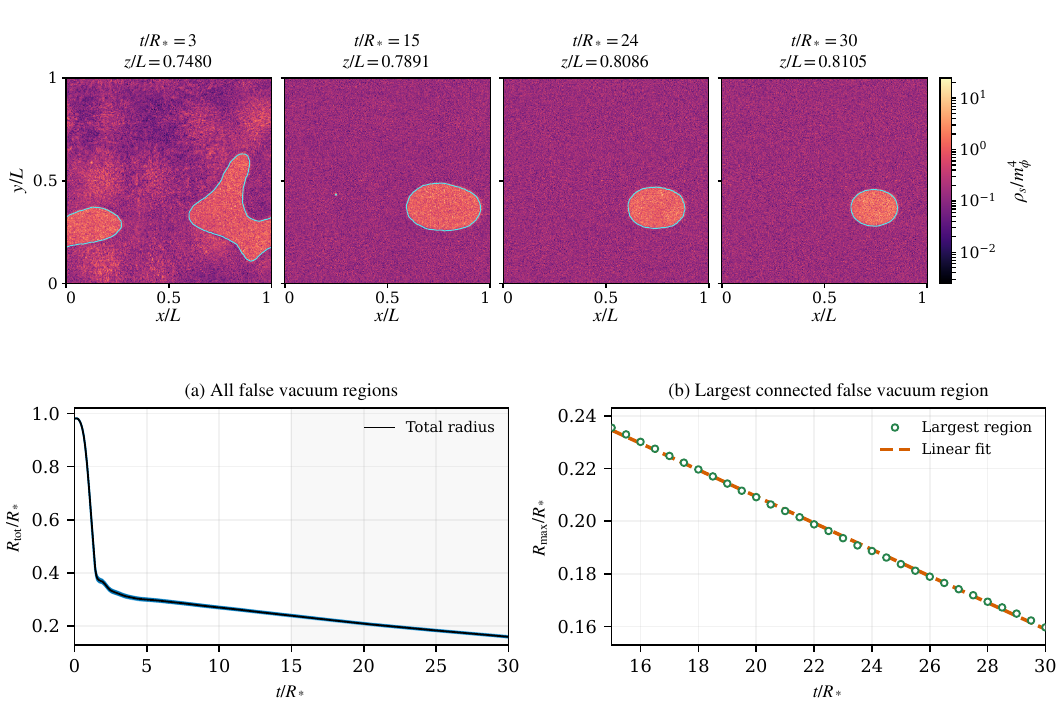}
    \caption{
    False vacuum regions in S$_L$ at late times. Top: \(\rho_s=\rho_{\mathrm K,s}+\rho_{\mathrm G,s}\) on the plane with the largest false vacuum area at each time. Contours mark \(\phi=\phi_{\rm true}/2\). Bottom: (a) radius \(R_{\rm tot}\), defined from the total false vacuum volume. The interval \(15\leq t/R_*\leq30\) is shaded. The blue band shows the variation in \(R_{\rm tot}\) when the threshold is changed from \(0.4\phi_{\rm true}\) to \(0.6\phi_{\rm true}\). (b) Radius \(R_{\rm max}\), defined from the volume of the largest connected region, and the dashed line showing its linear fit over the same interval. 
    }
    \label{fig:energy_density_evolution_3x4_smalllattice}
\end{figure*}

The extended S$_L$ run follows changes in the energy distribution beyond bubble merger. The spatial average over the periodic box is \(\langle X\rangle_V=L^{-3}\int_V d^3x\,X\). Fig.~\ref{fig:2_energy_all_components} shows the kinetic, gradient, and potential contributions separately, with \(U_\phi=\langle V(\phi,0)\rangle_V\), \(U_s=\langle V(0,s)\rangle_V\), and \(U_{\rm int}=\lambda_{\phi s}\langle\phi^2s^2\rangle_V/4\). The field energies excluding the interaction term are \(E_a^0=K_a+G_a+U_a\), with \(K_a=\langle\rho_{\mathrm K,a}\rangle_V\) and \(G_a=\langle\rho_{\mathrm G,a}\rangle_V\), for \(a=\phi,s\). These energy densities are normalized in the figure by the initial total energy density \(E_0=E_\phi^0(0)+E_s^0(0)+U_{\rm int}(0)\), with no shift in the potential.

As \(U_\phi\) decreases rapidly during expansion and merger, energy is transferred mainly to \(s\)-field excitations and the interaction term, with a smaller part remaining in \(\phi\) motion and gradients. Most of this transfer occurs before \(2R_*\). After roughly \(5R_*\), the direction of energy transfer changes: the \(s\)-field energy excluding the interaction decreases slowly, while the corresponding \(\phi\)-field energy and \(U_{\rm int}\) increase. At late times, \(K_s\) is nearly constant. The decrease in \(E_s^0\) therefore comes mainly from \(G_s\) and \(U_s\). Meanwhile, \(K_\phi\) and \(G_\phi\) grow as \(U_\phi\) continues to fall.
Over \(0\) to \(30R_*\), the net increases in \(E_s^0\) and \(U_{\rm int}\) are about \(67\%\) and \(31\%\) of the decrease in \(E_\phi^0\), respectively.

Fig.~\ref{fig:energy_density_evolution_3x4_smalllattice} shows \(s\)-field energy concentrated inside false vacuum regions long after merger. The volume with \(\phi<\phi_{\rm true}/2\) is denoted by \(V_{\rm FV}\). The ratio \(V_{\rm FV}/L^3\) decreases from \(1.44\%\) to \(0.426\%\) between \(15R_*\) and \(30R_*\). Fig.~\ref{fig:energy_density_evolution_3x4_smalllattice}(a) gives the radius defined from the total false vacuum volume, \(R_{\rm tot}=(3V_{\rm FV}/4\pi)^{1/3}\). Over this interval, the largest connected region contains over \(95\%\) of the false vacuum volume and therefore represents the dominant remnant. We identify connected regions using the six nearest lattice sites and periodic boundary conditions. For the largest region of volume \(V_{\rm max}\), the radius \(R_{\rm max}=(3V_{\rm max}/4\pi)^{1/3}\) decreases approximately linearly [Fig.~\ref{fig:energy_density_evolution_3x4_smalllattice}(b)],
\begin{equation}
 \frac{R_{\rm max}}{R_*}\simeq0.23468-0.005051\left(\frac{t}{R_*}-15\right).
 \label{eq:remnant_linear_fit}
\end{equation}
The fit applies to \(15\leq t/R_*\leq30\).

The increase in the effective mass can suppress the transmission of low-frequency \(s\) excitations into the true vacuum and may therefore provide internal support during contraction. This trapping and the approximately linear decrease of \(R_{\rm max}\) motivate a comparison with cooling thermal balls \cite{Kawana:2022lba,Li:2026yhv}. In the slow regime of surface cooling of Ref.~\cite{Kawana:2022lba}, energy loss reduces the internal pressure. The vacuum pressure then drives contraction. This contraction transfers energy to the trapped particles and maintains approximate pressure balance at nearly constant temperature. A nearly steady outward energy flux per unit area then gives an approximately constant contraction speed. By contrast, slow volumetric cooling gives approximately exponential contraction when the internal energy density and the energy loss per unit volume are nearly constant \cite{Kawana:2022lba}.

The radius \(R_{\rm max}\) decreases approximately linearly, as in the regime of surface cooling, but the internal evolution differs from its simplest thermal description. Averaged over lattice sites with \(\phi<\phi_{\rm true}/2\), the \(s\)-field energy density \(\rho_{\mathrm K,s}+\rho_{\mathrm G,s}+V(0,s)\) increases by about \(28\%\) between \(15R_*\) and \(30R_*\). However, the integral of this energy density over the same region decreases. This component therefore does not follow the limit of a relativistic gas with constant energy density and nearly constant temperature in Ref.~\cite{Kawana:2022lba}. The treatment of energy transfer also differs from that in Ref.~\cite{Li:2026yhv}. In our model, the massive real scalar field \(s\) has the self-interaction \(\lambda_s s^4/4\) and exchanges energy with \(\phi\) through their coupling. Unlike the dissipative evolution in that study, we impose no damping term.
Confinement and slow contraction are shared features. 
However, identifying a thermal ball additionally requires thermal pressure support.

Unlike the model with a complex scalar field in Ref.~\cite{Li:2026yhv}, the present model has no conserved internal \(U(1)\) charge to stabilize a Q-ball. With continued energy loss, eventual evaporation would be more likely. If the rate of energy loss becomes sufficiently small, the remnant could survive for a long time because of the pressure from trapped excitations, as in the picture of a thermal ball \cite{Kawana:2022lba}. Long-lived oscillatory configurations, such as oscillons, provide another possibility in real scalar theories \cite{Copeland:1995fq}. The final state of the simulated remnant therefore remains open.

\section{Gravitational waves}
\label{Sec:GW}

\subsection{Source and spectral normalization} \label{Subsec:GW from FOPT}
GWs are sourced by the transverse-traceless (TT) part of the spatial components of the energy-momentum tensor. For the two scalar fields, terms proportional to \(\delta_{ij}\) do not contribute. The tensor before TT projection can therefore be written as
\begin{equation}
T_{ij}(\mathbf x, t) = \partial_i\phi\,\partial_j\phi + \partial_i s\,\partial_j s.
\end{equation}
Following Ref.~\cite{Cutting:2018tjt}, we introduce an auxiliary tensor \(u_{ij}\). In the continuum, it satisfies the following equation, with \(\Box\equiv\partial_t^2-\nabla^2\):
\begin{equation}
\Box\, u_{ij} = 16\pi G\, T_{ij}.
\end{equation}
The continuum Fourier convention is
\begin{equation}
\tilde T_{ij}({\mathbf k},t)=\int d^3x\, e^{-i{\mathbf k}\cdot{\mathbf x}}\,
T_{ij}({\bf x},t).
\label{eq:stress-fourier}
\end{equation}
On the lattice, the spatial integral in Eq.~\eqref{eq:stress-fourier} is evaluated as a sum with volume element \((\Delta x)^3\). Let \(k_{\rm eff}^2(\mathbf k)\) denote the eigenvalue of the negative lattice Laplacian. The Fourier coefficients of the auxiliary tensor \(u_{ij}\) then satisfy
\begin{equation}
\ddot{ u}_{ij}(\mathbf k, t) + k^2_{\rm eff}(\mathbf k)\,  u_{ij}(\mathbf k, t) = 16\pi G\, \tilde T_{ij}(\mathbf k, t).
\end{equation}
A leapfrog integrator evolves this tensor with the same time step as for \(\phi\) and \(s\). 
We use the 19-point Laplacian, whose leading discretization error is isotropic \cite{https://doi.org/10.1002/num.20129,Frolov:2008hy}. The explicit expression for \(k_{\rm eff}^2\) is
\begin{equation}
\label{eq:k-eff-19}
\begin{aligned}
    k _ {\mathrm {e f f}} ^ {2} (\mathbf {k}) 
    &= \frac {1}{\Delta x ^ {2}} \left[ 4 - \frac {2}{3} \sum_ {\mu} \cos \left(k _ {\mu} \Delta x\right) \right. \\
    & \hspace{4em} \left. - \frac {2}{3} \sum_ {\mu < \nu} \cos \left(k _ {\mu} \Delta x\right) \cos \left(k _ {\nu} \Delta x\right) \right]
\end{aligned}
\end{equation}
where \(\Delta x\) is the lattice spacing and \(k_\mu=2\pi n_\mu/L\) in a periodic box of side \(L\). The transverse-traceless (TT) projection is applied when evaluating the spectrum,
\begin{equation}
 h^{\rm TT}_{ij}(\mathbf k, t) = \Lambda_{ij,lm}(\hat{\mathbf k})\, u_{lm}(\mathbf k, t),
\end{equation}
with the TT projector
\begin{equation}
\Lambda_{ij,lm}(\hat{\mathbf k}) = P_{il}(\hat{\mathbf k})P_{jm}(\hat{\mathbf k}) - \tfrac{1}{2}P_{ij}(\hat{\mathbf k})P_{lm}(\hat{\mathbf k}), 
\end{equation}
\begin{equation}
P_{ij}(\hat{\mathbf k}) = \delta_{ij} - \hat k_i\hat k_j,
\end{equation}
In the continuum, \(\hat{\mathbf k}=\mathbf k/k\). In the simulations, the TT projection uses \(\hat{\mathbf q}\) in place of \(\hat{\mathbf k}\), where \(q_i=\sin(k_i\Delta x/2)\) and \(\hat q_i=q_i/(\sum_j q_j^2)^{1/2}\). The resulting tensor is transverse to \(\mathbf q\). The zero mode is excluded from the projection. The choice of lattice derivatives affects this projection, as discussed in Ref.~\cite{Figueroa:2011ye}. For comparison with the continuum normalization, define
\begin{equation}
\left\langle\dot{h}_{i j}^{\mathrm{TT}}(\mathbf{k}, t) \dot{h}_{i j}^{\mathrm{TT}}\left(\mathbf{k}^{\prime}, t\right)\right\rangle=P_{\dot{h}}(\mathbf{k}, t)(2 \pi)^{3} \delta\left(\mathbf{k}+\mathbf{k}^{\prime}\right) ,
\end{equation}
so that the power spectrum of the GW energy density, normalized to \(\rho_c\), is
\begin{equation}
\frac{d\Omega_{\rm gw}}{d\ln k}(\mathbf{k}, t) = \frac{1}{32\pi G\rho_c}\,\frac{k^3}{2\pi^2}\,P_{\dot h}(\mathbf{k}, t).
\end{equation}
Here \(\rho_c=3H_*^2/(8\pi G)\), \(\Omega_{\rm vac}=\rho_{\rm vac}/\rho_c\), and \(R_*\) is defined in Sec.~\ref{Sec:multi}. The spectra are divided by \((H_*R_*\Omega_{\rm vac})^2\) to remove their leading dependence on these scales. The Hubble rate enters only this normalization: we neglect cosmological expansion, assuming simulation lengths and times much smaller than the Hubble scale.

To calculate the contributions from \(\phi\) and \(s\) separately, we evolve an auxiliary tensor for each field. Each tensor is sourced by the gradient terms of the corresponding field. Their amplitudes are added before calculating the total power. The normalized spectra are denoted by \(\widetilde\Omega_{\rm gw}^{(a)}\equiv(H_*R_*\Omega_{\rm vac})^{-2}d\Omega_{\rm gw}^{(a)}/d\ln k\), with \(a=\phi,s,\mathrm{tot},\mathrm{cross}\). The total spectrum is \(\widetilde\Omega_{\rm gw}^{(\mathrm{tot})}=\widetilde\Omega_{\rm gw}^{(\phi)}+\widetilde\Omega_{\rm gw}^{(s)}+\widetilde\Omega_{\rm gw}^{(\mathrm{cross})}\), where the cross term can have either sign. This separates contributions from the two fields, but does not distinguish bubble walls, energy shells and oscillations after collision within each field. The spectra are averaged over spherical shells in momentum space throughout the evolution.

\begin{figure*}[t]
    \centering
    \includegraphics[width=0.98\textwidth]{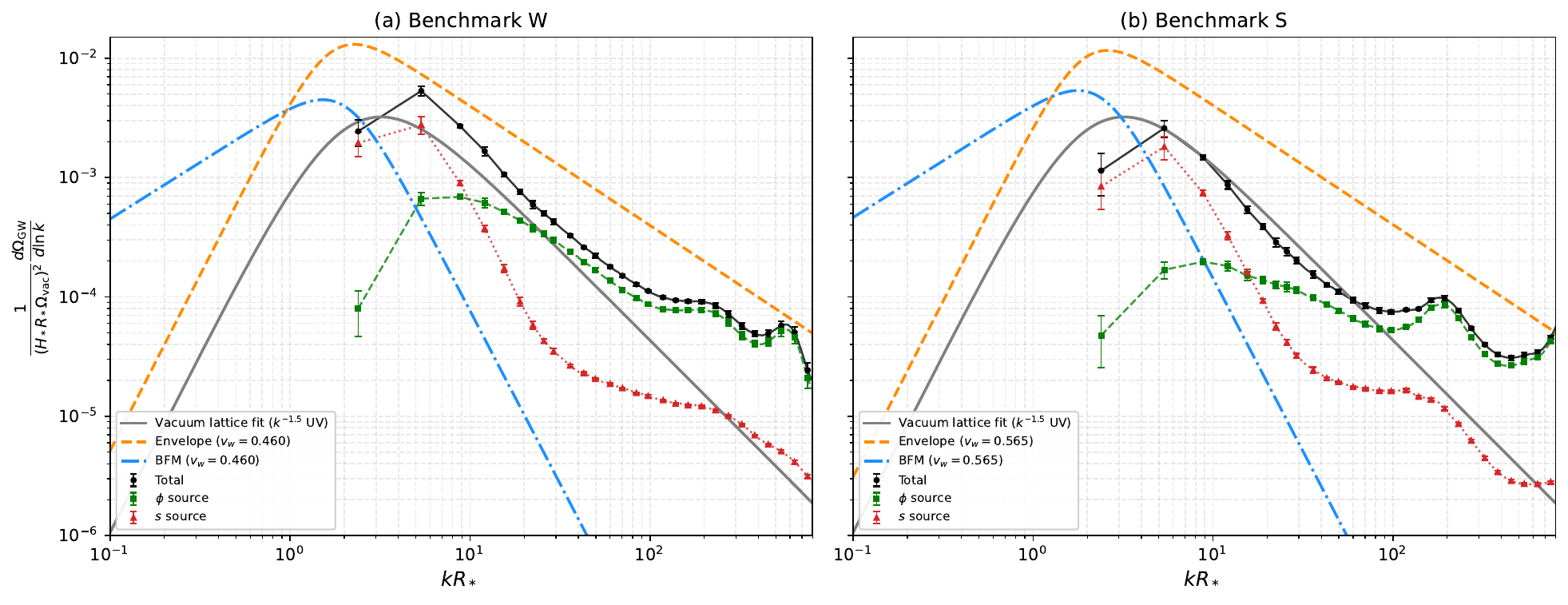}
    \caption{
    Mean gravitational-wave spectra at \(t/R_*=5.00\) for W (a) and S (b), each averaged over five simulations with different random initial conditions. Error bars show the sample standard deviation across these simulations. The black solid, green dashed, and red dotted curves show \(\widetilde\Omega_{\rm gw}^{(\mathrm{tot})}\), \(\widetilde\Omega_{\rm gw}^{(\phi)}\), and \(\widetilde\Omega_{\rm gw}^{(s)}\). The total also includes the term denoted by \(\mathrm{cross}\). The grey, orange dashed, and blue dash-dotted curves are the vacuum lattice fit \cite{Cutting:2018tjt}, envelope fit, and BFM fit \cite{Konstandin:2017sat}. 
    The envelope and BFM reference curves use the wall velocities \(v_w=0.460\) for W and \(0.565\) for S. The shape parameters are those for small wall velocities, and the peak quantities follow the velocity-dependent fits in Ref.~\cite{Konstandin:2017sat}. These curves are approximate references at the simulated wall speeds.
    Modes with \(kR_*\geq800\) are excluded.
                }

    \label{fig:gw-spectra}
\end{figure*}

\subsection{Gravitational-wave spectra} \label{Subsec: Power Spectra of Gravitational Wave}

Fig.~\ref{fig:gw-spectra} shows the mean spectra at \(t/R_*=5.00\). Both peak at the sampled wave number \(kR_*=5.36\), with normalized amplitudes \((5.32\pm0.52)\times10^{-3}\) for W and \((2.59\pm0.42)\times10^{-3}\) for S. The uncertainties are sample standard deviations at that bin. At the peak, \(\widetilde\Omega_{\rm gw}^{(s)}\) exceeds \(\widetilde\Omega_{\rm gw}^{(\phi)}\) and contributes \(0.52\) and \(0.70\) of the total in W and S. The cross term contributes \(0.35\) and \(0.24\) of the total, respectively.

Over \(8\leq kR_*\leq20\), the fitted slopes of \(\widetilde\Omega_{\rm gw}^{(s)}\) are \(-2.98\pm0.10\) for W and \(-2.72\pm0.06\) for S. The BFM reference curves have slopes of \(-2.90\) and \(-2.89\). The total spectra are shallower, with slopes \(-1.66\pm0.05\) and \(-1.76\pm0.05\), because they also include the \(\phi\) contribution and the cross term.

The similar slopes motivate a comparison between the energy shells in \(s\) and the BFM source. Appendix~\ref{app:A} presents this comparison. For a thin shell with predominantly radial gradients, the stress tensor is proportional to \(\hat\xi_i\hat\xi_j\). When the energy stored in the shell per unit solid angle grows as \(R^3\), its energy per unit area grows as \(R\). Let \(\sigma_s=\int dr\,(\partial_r s)^2\), where the integral is taken across the shell. If the ratio of \(\sigma_s\) to the energy per unit area remains approximately constant, then \(\sigma_s\propto R\). Before collision, its Fourier transform then has the same \(R^3\) dependence and angular TT structure as the BFM, apart from the overall normalization \cite{Konstandin:2017sat}. Although the spectral slope also depends on the correlations of the source at different times, this common geometry provides a possible basis for the similar slopes over \(8\leq kR_*\leq20\).

After collision, the shells in the BFM continue to propagate, with energy per unit area proportional to \(R^{-2}\). The energy shells in \(s\) can continue to propagate after the first collision, but become distorted as energy spreads in W and remains concentrated in localized regions in S. This differs from the continued propagation assumed in the BFM. At high wave number, finite shell thickness and gradients perpendicular to the radial direction also limit the description by a thin shell. In Fourier space, this approximation is valid only for \(kd_{\rm sh}\ll1\). These effects may explain why the spectrum differs from the BFM reference curve in the high-wave-number (UV) direction.

Beyond the peak, \(\widetilde\Omega_{\rm gw}^{(\phi)}\) first exceeds \(\widetilde\Omega_{\rm gw}^{(s)}\) at the sampled values \(kR_*=12.1\) in W and \(19.0\) in S. Over \(80\leq kR_*\leq250\), the total spectrum decreases more slowly in W and rises slightly in S. Both spectra decrease more steeply over \(250\leq kR_*\leq500\), unlike the single power-law tails of the envelope and BFM references. This broad feature in the high-wave-number (UV) direction is less pronounced than the secondary peak from oscillations after collision in simulations of vacuum transitions with a single scalar field \cite{Cutting:2018tjt,Cutting:2020nla}.
Table~\ref{tab:spectral-slopes} shows the slopes of the total spectrum over \(20\leq kR_*\leq500\). These fits over finite intervals describe the changing shapes of the total spectra, and do not establish a separate power law in each asymptotic regime.
\begin{table}[!t]
\caption{Fitted slopes \(n\) of the total GW spectrum, \(\widetilde\Omega_{\rm gw}^{(\mathrm{tot})}\propto k^n\), over the listed intervals. All slopes quoted for the simulated W and S spectra are means of separate fits to the five simulations in each set, with sample standard deviations.}
\label{tab:spectral-slopes}
\begin{ruledtabular}
\begin{tabular}{ccc}
\(kR_*\) interval & W & S \\
\hline
\(20\)--\(80\) & \(-1.19\pm0.03\) & \(-1.02\pm0.03\) \\
\(80\)--\(250\) & \(-0.33\pm0.04\) & \(+0.09\pm0.05\) \\
\(250\)--\(500\) & \(-0.64\pm0.09\) & \(-1.01\pm0.11\)
\end{tabular}
\end{ruledtabular}
\end{table}

The envelope and BFM fits have ultraviolet tails \(k^{-1}\) and \(k^{-2.9}\), respectively, using the shape parameters for small wall velocities of Ref.~\cite{Konstandin:2017sat}. In simulations of vacuum transitions with a single scalar field, the fitted collision spectrum falls approximately as \(k^{-1.5}\) in Ref.~\cite{Cutting:2018tjt}. Varying the potential gives slopes from about \(-1.4\) to \(-2.2\) in Ref.~\cite{Cutting:2020nla}. The latter fits exclude the oscillation bump at high wave number. The reference curves use \(\beta R_*=(8\pi)^{1/3}v_w\), the relation for exponential nucleation at constant wall speed, to express their scales in units of \(R_*\) \cite{Cutting:2018tjt}. We set their efficiency to \(\kappa=1\). This places all available vacuum energy in the assumed source. The amplitude scales as \(\kappa^2\).

Near \(kR_*=180\), the ratios \(\widetilde\Omega_{\rm gw}^{(\phi)}/\widetilde\Omega_{\rm gw}^{(\mathrm{tot})}\) are about \(0.85\) in W and \(0.87\) in S. Simulations of vacuum transitions with a single scalar field show a UV feature produced by field oscillations after collision \cite{Cutting:2018tjt,Cutting:2020nla}. In Ref.~\cite{Cutting:2018tjt}, this feature is associated with the scalar mass scale and the initial wall thickness before Lorentz contraction. Since \(\phi\) provides most of the power in this range, field oscillations after collision may also contribute to the UV feature in our simulations. Energy transfer between \(\phi\) and \(s\) can change the subsequent field evolution and may also affect this broad feature in the GW spectrum. The separate contributions of these processes have not been determined. The feature also appears in W, so its presence alone does not identify a contribution from the long-lived false vacuum remnants in S.

The remnants are regions of false vacuum surrounded by the true vacuum. Energy in \(s\) remains concentrated inside them, while gradients of \(\phi\) are present at their boundaries. Changes in shape and oscillations that break spherical symmetry can produce a time-dependent TT source and generate GWs. Any GW signal from these remnants could overlap the radiation from bubble collisions and later field oscillations.  Slow, nearly spherical evolution would further suppress radiation. The stress in each field includes contributions from both the remnants and other field configurations. The separate field spectra therefore cannot identify the contribution from the remnants alone. The total spectrum also includes interference between the two field contributions. Thus, these spectra do not establish a distinct GW signal from the long-lived remnants.

\section{Conclusions and discussion}
\label{Sec:Conclusion and Discussion}
 
We have studied gravitational-wave production in a model where the scalar order parameter \(\phi\) is coupled to an additional scalar field \(s\). The field \(s\) is initialized with fluctuations. Starting from the initial bubbles, both fields evolve without an imposed fluid profile or a phenomenological friction term. Energy accumulates near the expanding walls and affects their motion. During collisions, the shells lose their original shape and their energy distribution changes, unlike the freely propagating sound shells of an acoustic description.
For the larger coupling, false vacuum remnants persist at least until \(t/R_*=30\), while their total volume decreases by about \(70\%\) over \(15\leq t/R_*\leq30\). The radius defined from the volume of the largest remnant decreases approximately linearly. An approximately linear contraction also occurs during surface cooling of a thermal ball \cite{Kawana:2022lba}. In that model, the temperature and energy density remain approximately constant. By contrast, the mean \(s\)-field energy density inside our false vacuum regions increases over the same interval. The similar contraction alone therefore does not establish that these remnants are thermal balls. Continued energy loss may eventually cause the remnants to disappear. If the rate of energy loss decreases, pressure from trapped excitations could allow them to survive for much longer. Other long-lived field configurations may also form. Their final fate has therefore not been determined.

The GW spectra peak at \(kR_*\simeq5.4\), where the contribution from \(s\) exceeds that from \(\phi\). Over \(8\leq kR_*\leq20\), the spectrum from \(s\) has a slope similar to that of the BFM reference curve. Under the assumptions of Appendix~\ref{app:A}, the \(s\)-field stress before collision can be approximated by a thin shell source. Its Fourier integrand has the same dependence on radius and the same angular structure as the BFM, apart from the overall normalization. After collision, changes in the shape of the \(s\)-field shells and their interaction with \(\phi\) lead to a different evolution of the source. These differences may explain the departure of the \(s\)-field spectrum from the BFM reference curve in the high-wave-number (UV) direction. In the high-wave-number (UV) direction, the \(\phi\) contribution dominates and the total spectrum develops a broad feature. This feature is less pronounced than the secondary peak found in simulations of vacuum transitions with a single scalar field \cite{Cutting:2018tjt,Cutting:2020nla}. The feature appears in both W and S, so its presence alone does not identify a contribution from the long-lived false vacuum remnants in S.

After bubble merger, part of the released energy remains in the long-lived remnants. This motivates extending GW source models based on expanding shells to include localized stresses with a different lifetime and geometry. These remnants are also relevant to models of compact relics and primordial black holes from trapped particles \cite{Kawana:2022lba,Baker:2021nyl}. In these models, the outcome depends on the pressure of the trapped particles, cooling and gravitational collapse. The observed contraction is relevant to these scenarios, but does not establish either final state. 

The peak of the total GW spectrum and its shape at high frequencies provide information about the transition dynamics. These features could help constrain the transition parameters and distinguish competing source models \cite{Hindmarsh:2017gnf,Caprini:2019egz}. This requires a sufficiently strong signal that is redshifted into the sensitivity band of a space-based interferometer such as LISA.

\vspace{5pt}
\noindent
\section*{Acknowledgments}
This work is supported in part by the National Natural Science Foundation of China under Grants No. 12235019 and No. 12475067.

\appendix

\section{Thin shell description and its limitations}\label{app:A}

We calculate the Fourier transform and TT projection of a thin shell of the \(s\) field, compare this source with the BFM before collision, and discuss how the evolution after collision affects the source correlations.

\subsection{Stress tensor of a thin scalar shell}
For the real scalar field \(s\), the gradient part of the spatial stress is
\begin{equation}
T^{(s)}_{ij}(\mathbf{x},t)=\partial_i s\,\partial_j s.
\label{eq:app_micro_stress}
\end{equation}
Terms proportional to \(\delta_{ij}\) vanish under TT projection. We consider a spherical shell of radius \(R_n\) and width \(d_{\rm sh}\ll R_n\), with predominantly radial gradients,
\begin{equation}
|\partial_r s|^2\gg |\boldsymbol\nabla_{\!\parallel}s|^2.
\label{eq:app_radial_dom}
\end{equation}
Here \(\partial_r=\hat{\xi}\cdot\boldsymbol\nabla\), \(\hat{\xi}=(\mathbf{x}-\mathbf{x}_n)/|\mathbf{x}-\mathbf{x}_n|\), and \(\boldsymbol\nabla_{\!\parallel}s=\boldsymbol\nabla s-\hat{\xi}\,\partial_r s\). The shell structure in the energy density does not by itself establish Eq.~\eqref{eq:app_radial_dom}. The leading radial stress is
\begin{equation}
T^{(s)}_{ij}\simeq(\partial_r s)^2\hat{\xi}_i\hat{\xi}_j.
\label{eq:app_radial_stress}
\end{equation}

Let \(\mathbf{x}_n\) and \(t_n\) denote the centre and effective nucleation time of bubble \(n\). For \(t\geq t_n\), define
\begin{equation}
\tau_n\equiv t-t_n\geq0,
\label{eq:app_coh_time}
\end{equation}
and adopt the reference trajectory with constant wall speed
\begin{equation}
R_n(t)=v_w\tau_n.
\label{eq:app_coh_radius}
\end{equation}
The source vanishes for \(t<t_n\). This trajectory neglects the finite initial radius and acceleration of the simulated bubbles. The time of the first collision \(t_c^{(n)}(\hat{\xi})>t_n\) marks the change in the source weight below.

Before collision, if \(f(r)\) varies slowly across the shell, the integrated stress can be approximated by
\begin{equation}
\int_{R_n-d_{\rm sh}/2}^{R_n+d_{\rm sh}/2}dr\,f(r)(\partial_r s)^2
\simeq f(R_n)\,\sigma_s^{(n)}(\hat{\xi},t),
\label{eq:app_surface_limit}
\end{equation}
where
\begin{equation}
        \sigma_s^{(n)}(\hat{\xi},t)
        \equiv
        \int_{R_n-d_{\rm sh}/2}^{R_n+d_{\rm sh}/2}
        dr\,(\partial_r s)^2
        \label{eq:app_sigma_def}
\end{equation}
is the coefficient of the shell contribution in Eq.~\eqref{eq:app_coherent_source}. In particular, \(\sigma_s^{(n)}=2\int dr\,\rho_{\mathrm G,r}\), where \(\rho_{\mathrm G,r}=(\partial_r s)^2/2\). If \(\partial_r s=0\) outside this interval, the error in the integral is bounded by \(\sigma_s^{(n)}d_{\rm sh}\sup|f'|/2\). In a spatial Fourier transform, the radial measure and phase therefore require both \(d_{\rm sh}/R_n\ll1\) and \(kd_{\rm sh}\ll1\). The thin shell contribution is written in terms of \({\cal D}_n\) and \(\sigma_{s,{\rm ref}}^{(n)}\) below:
\begin{equation}
\begin{aligned}
    T^{(s),{\rm shell}}_{ij}(\mathbf{x},t)
    &=
    \sum_n
    {\cal D}_n(\hat{\xi},t)\,
    \sigma_{s,{\rm ref}}^{(n)}(t)\,
    \hat{\xi}_i\hat{\xi}_j\, \\
    & \hspace{3em} \delta\!\left(|\mathbf{x}-\mathbf{x}_n|-R_n(t)\right).
\end{aligned}
\label{eq:app_coherent_source}
\end{equation}
The factor \({\cal D}_n(\hat{\xi},t)\) is the non-negative effective weight of this radial stress. Before the first collision,
\begin{equation}
        {\cal D}_n(\hat{\xi},t)=1,
        \quad t_n\leq t<t_c^{(n)}(\hat{\xi}).
        \label{eq:app_sharp_D}
\end{equation}
Its subsequent evolution is not fixed here. In our simulations, the energy shells in \(s\) continue to propagate for some distance after collision.
For this comparison, \(dE_s^{(n)}\) is the kinetic and gradient energy of \(s\) in the shell element.
When the energy stored in a shell element of solid angle \(d\Omega\) is proportional to \(R_n^3\), we write
\begin{equation}
        dE_s^{(n)}(\hat{\xi},t)
        \simeq
        \frac{\rho_0}{3}
        \left[R_n(t)\right]^3 d\Omega ,
        \quad t<t_c^{(n)}(\hat{\xi}),
        \label{eq:app_swept_energy}
\end{equation}
where \(\rho_0\) is a constant that sets the energy in this relation. For a finite initial radius, \(R_n^3\) is replaced by \(R_n^3-R_0^3\) if the initial shell energy is neglected. For comparison with the simulated bubbles, neglecting this correction requires \(R_n\gg R_0\). The dimensionless ratio \(q_s\) compares \(\sigma_s^{(n)}\) with the shell energy per unit area. In this reference model, we assume the same approximately constant \(q_s\) for all surface elements before their first collision. This gives
\begin{equation}
\begin{aligned}
        \sigma_s^{(n)}(\hat{\xi},t)
        &\simeq \sigma_{s,{\rm ref}}^{(n)}(t) \\
        &\equiv \frac{q_s\rho_0}{3}R_n(t)
        =\frac{q_s\rho_0 v_w}{3}\tau_n.
\end{aligned}
        \label{eq:app_sigma_scaling}
\end{equation}
Here \(q_s=\sigma_s^{(n)}R_n^2/(dE_s^{(n)}/d\Omega)\). After collision, the stress is parameterized by \({\cal D}_n\sigma_{s,{\rm ref}}^{(n)}\), with \({\cal D}_n\) to be fitted to the simulations.

The TT source in Fourier space of an exactly spherical radial shell vanishes. Angular dependence of \({\cal D}_n\), for example through the time of the first collision, can break that symmetry. Real field configurations can also contain angular structure before collision. If the radial contribution vanishes under TT projection, the terms neglected in Eq.~\eqref{eq:app_radial_stress} can provide the leading GW source.

\subsection{Fourier transform}
The Fourier transform over the time interval ending at \(t_f\) is defined by
\begin{equation}
\begin{aligned}
        \widetilde T^{(s)}_{ij}(\mathbf{k},\omega;t_f)
        &=\int_{-\infty}^{t_f}dt\int d^3\mathbf{x}\,\\
        &\quad\times e^{i\omega t-i \mathbf{k}\cdot \mathbf{x}}
        T^{(s),{\rm shell}}_{ij}(\mathbf{x},t).
\end{aligned}
        \label{eq:app_FT_def}
\end{equation}
The dependence on \(t_f\) is suppressed below. The upper limit \(t_f\) sets the time interval of the Fourier integral, rather than the lifetime of the source. Substitution of Eq.~\eqref{eq:app_coherent_source}, with \(\mathbf{x}=\mathbf{x}_n+r\hat{\xi}\), gives
\begin{equation}
\begin{aligned}
&\widetilde T^{(s)}_{ij}(\mathbf{k},\omega)
=
\sum_{n:t_n<t_f} e^{-i\mathbf{k}\cdot \mathbf{x}_n} \\
&\quad\times\int_{t_n}^{t_f}dt\,e^{i\omega t}
\int d\Omega_{\hat{\xi}}\, \\
& \hspace{1em} \left[R_n(t)\right]^2
{\cal D}_n(\hat{\xi},t)
\sigma_{s,{\rm ref}}^{(n)}(t)
e^{-i\mathbf{k}\cdot\hat{\xi}R_n(t)}
\hat{\xi}_i\hat{\xi}_j .
\end{aligned}
\label{eq:app_FT_shell}
\end{equation}

Choose the wave vector along the \(z\)-axis, \(\mathbf{k}=k\hat z\), and write
\begin{equation}
        \hat{\xi}
        =
        \left(
        \sqrt{1-\zeta^2}\cos\varphi,\,
        \sqrt{1-\zeta^2}\sin\varphi,\,
        \zeta
        \right),
        \quad
        \zeta=\cos\theta .
        \label{eq:app_xi_param}
\end{equation}
Here \(\varphi\) is the azimuthal angle about \(\hat{\mathbf{k}}\), so that \(\mathbf{k}\cdot\hat{\xi}=k\zeta\). Substituting Eq.~\eqref{eq:app_sigma_scaling} into Eq.~\eqref{eq:app_FT_shell} gives
\begin{equation}
\begin{aligned}
&\widetilde T^{(s)}_{ij}(k,\omega)
 =
\frac{q_s\rho_0 v_w^3}{3}
\sum_{n:t_n<t_f} e^{-i\mathbf{k}\cdot \mathbf{x}_n} \\
&\quad\times\int_{t_n}^{t_f}dt\,e^{i\omega t}
\int_{-1}^{1}d\zeta
\int_0^{2\pi}d\varphi\, \\
& \hspace{3em} {\cal D}_n(\zeta,\varphi,t)
\left[\tau_n\right]^3
e^{-iv_w k\tau_n\zeta}
\hat{\xi}_i\hat{\xi}_j .
\end{aligned}
\label{eq:app_FT_final}
\end{equation}

\subsection{Transverse-traceless projection}

For \( \mathbf{k}=k\hat z\), the transverse projector and the TT projector for symmetric tensors are
\begin{equation}
\begin{aligned}
    &P_{ij}=\delta_{ij}-\hat z_i\hat z_j
    ={\rm diag}(1,1,0)_{ij}, \\
    &\Lambda_{ij,lm}=P_{il}P_{jm}
    -\frac12 P_{ij}P_{lm}.
\end{aligned}
\label{eq:app_TT_projector}
\end{equation}
Applying this projector to the radial tensor \(\hat{\xi}_l\hat{\xi}_m\) gives
\begin{equation}
\Lambda_{ij,lm}\hat{\xi}_l\hat{\xi}_m
 =
\frac{1-\zeta^2}{2}
\left[
\cos(2\varphi)e^+_{ij}
+
\sin(2\varphi)e^\times_{ij}
\right],
\label{eq:app_TT_xixi}
\end{equation}
where the two polarization tensors are
\begin{equation}
        e^+_{ij}={\rm diag}(1,-1,0)_{ij},
        \quad
        e^\times_{ij}=
        \begin{pmatrix}
        0&1&0\\
        1&0&0\\
        0&0&0
        \end{pmatrix}_{ij}.
        \label{eq:app_pol_basis}
\end{equation}
For this choice of direction, the dependence of the TT projection on the azimuthal angle is given by \(\cos(2\varphi)\) and \(\sin(2\varphi)\). If the remaining integrand is independent of \(\varphi\), both angular integrals vanish. A spherical shell therefore has no TT source.

The source after TT projection can be written as
\begin{equation}
\Lambda_{ij,lm}\widetilde T^{(s)}_{lm}
=
\frac12
\left[
C^{(s)}_+(k,\omega)e^+_{ij}
+
C^{(s)}_\times(k,\omega)e^\times_{ij}
\right],
\label{eq:app_C_def}
\end{equation}
where \(C_p^{(s)}=e^p_{ij}\widetilde T^{(s)}_{ij}\). Since \(e^p_{ij}e^{p'}_{ij}=2\delta_{pp'}\), the squared magnitude of the source after TT projection is \((|C_+^{(s)}|^2+|C_\times^{(s)}|^2)/2\). This normalization is used throughout this appendix. The coefficients in Ref.~\cite{Konstandin:2017sat} use a different convention. Explicitly,
\begin{equation}
\begin{aligned}
&C^{(s)}_p(k,\omega)
 =
\frac{q_s\rho_0 v_w^3}{3}
\sum_{n:t_n<t_f} e^{-i\mathbf{k}\cdot \mathbf{x}_n} \\
&\quad\times\int_{t_n}^{t_f}dt\,e^{i\omega t}
\int_{-1}^{1}d\zeta\,
(1-\zeta^2)\, \\
& \int_0^{2\pi}d\varphi\, 
  w_p(\varphi)\,
{\cal D}_n(\zeta,\varphi,t)
\left[\tau_n\right]^3
\\ &\hspace{2em}e^{-iv_w k\tau_n\zeta}.
\end{aligned}
\label{eq:app_Cp}
\end{equation}
Here
\begin{equation}
        w_+(\varphi)=\cos(2\varphi),
        \quad
        w_\times(\varphi)=\sin(2\varphi).
        \label{eq:app_wp}
\end{equation}
Eq.~\eqref{eq:app_Cp} separates the angular factors for a shell with radial gradients. The phase, the radius \(R_n\), and the factor \({\cal D}_n\) depend on time. Their evolution determines the frequency dependence of the source.

\subsection{Relation to the BFM}
We now compare the time dependence of the source with the BFM of Ref.~\cite{Konstandin:2017sat}. Let
\begin{equation}
\tau_c^{(n)}(\zeta,\varphi)\equiv t_c^{(n)}(\zeta,\varphi)-t_n>0.
\label{eq:app_tau_c}
\end{equation}
For continuum radiation we now impose \(\omega=k>0\). After the shift \(t=t_n+\tau_n\), the phase becomes
\begin{equation}
\begin{aligned}
e^{i\omega t-i\mathbf{k}\cdot\mathbf{x}_n}
&=e^{i\omega t_n-i\mathbf{k}\cdot\mathbf{x}_n}e^{i\omega\tau_n}\\
&=e^{i\omega(t_n-\hat{\mathbf{k}}\cdot\mathbf{x}_n)}e^{i\omega\tau_n}.
\end{aligned}
\label{eq:app_phase_shift}
\end{equation}
The on-shell condition is used only in this continuum comparison. The simulation evolves waves with the lattice dispersion. Eq.~\eqref{eq:app_Cp} gives
\begin{equation}
\begin{aligned}
C_p^{(s)}(k,\omega)
&=\frac{q_s\rho_0v_w^3}{3}\sum_{n:t_n<t_f} e^{i\omega(t_n-\hat{\mathbf{k}}\cdot\mathbf{x}_n)}\\
&\quad\times\int_{-1}^{1}d\zeta\,(1-\zeta^2)\int_0^{2\pi}d\varphi\,w_p(\varphi)\\
&\quad\times\int_0^{t_f-t_n}d\tau_n\,{\cal D}_n(\zeta,\varphi,t_n+\tau_n)\\
&\hspace{3em}\times\tau_n^3 e^{i\omega\tau_n(1-v_w\zeta)}.
\end{aligned}
\label{eq:app_precollision_match}
\end{equation}

In the BFM, \({\cal D}_n=1\) before the first collision and \({\cal D}_n=(\tau_c^{(n)}/\tau_n)^3\) afterwards.
For a direct comparison, combining Eqs.~(8)--(12) of Ref.~\cite{Konstandin:2017sat}, with the same final time \(t_f\), gives
\begin{equation}
\begin{aligned}
C_p^{\rm BFM}(\omega;t_f)
&=\frac{1}{12\pi}\sum_{n:t_n<t_f}e^{i\omega(t_n-\hat{\mathbf{k}}\cdot\mathbf{x}_n)}\\
&\quad\times\int_{-1}^{1}d\zeta\,(1-\zeta^2)\int_0^{2\pi}d\varphi\,w_p(\varphi)\\
&\quad\times\int_0^{t_f-t_n}d\tau_n\,\min(\tau_n,\tau_c^{(n)})^3\\
&\hspace{3em}\times e^{i\omega\tau_n(1-v_w\zeta)}.
\end{aligned}
\label{eq:app_BFM_Cp}
\end{equation}
In that convention, Eqs.~(4) and (7) of Ref.~\cite{Konstandin:2017sat} give the dimensionless GW spectral function
\begin{equation}
\Delta_{\rm BFM}(\omega;t_f)
=\frac{3v_w^6\beta^2\omega^3}{2\pi V}
\int d\hat{\mathbf{k}}\,\sum_{p=+,\times}|C_p^{\rm BFM}|^2,
\label{eq:app_BFM_spectrum}
\end{equation}
where \(V\) is the volume and \(\beta\) the inverse transition timescale in the BFM reference. The finite upper limit is used to compare the source integrals over the same time interval. To obtain the full BFM spectrum, the time integral must extend beyond \(t_f\), unless the contribution at later times is negligible. Before the first collision, \({\cal D}_n=1\) and \(\min(\tau_n,\tau_c^{(n)})^3=\tau_n^3\). The comparison uses the same bubble geometry and wall speed in Eqs.~\eqref{eq:app_precollision_match} and \eqref{eq:app_BFM_Cp}. Their integrands then differ only by the constant factor \(4\pi q_s\rho_0v_w^3\). Collisions break the spherical symmetry of the shells and can produce a nonzero TT source. This source depends on the shell structure before collision. If the shells also evolve in a similar way during collision, their common structure may help explain the similar spectral slopes over \(8\leq kR_*\leq20\). The spectrum also depends on the correlations of the source at different times.

After collision, the spherical shells in the BFM continue to propagate at constant speed, with energy per unit area proportional to \(R_n^{-2}\). In our simulations, changes in shell shape and in the energy distribution instead change the source geometry and its correlations at different times. Finite thickness and gradients perpendicular to the radial direction further limit this description. In particular, modes with \(kd_{\rm sh}\gtrsim1\) resolve the shell profile. These effects may explain the different spectra at larger \(kR_*\). The evolution of \({\cal D}_n\) after collision is not fixed here. This weight changes the amplitude of the radial stress, but does not describe changes in the direction of the gradients.

\bibliographystyle{apsrev4-2}
\bibliography{main,reference_exception_pending}

\end{document}